\documentclass[10pt,english,pre,nofootinbib,reprint,showpacs,showkeys,preprintnumbers,floatfix]{revtex4-1}
\usepackage[latin9]{inputenc}
\usepackage[a4paper]{geometry}
\usepackage{color}
\usepackage{babel}
\usepackage{units}
\usepackage{amsmath}
\usepackage{amssymb}
\usepackage{graphicx}
\usepackage{esint}
\usepackage[unicode=true,pdfusetitle,
 bookmarks=true,bookmarksnumbered=false,bookmarksopen=false,
 breaklinks=false,pdfborder={0 0 1},backref=false,colorlinks=false]
 {hyperref}

\makeatletter
 
 \@ifundefined{textcolor}{}
 {%
   \definecolor{BLACK}{gray}{0}
   \definecolor{WHITE}{gray}{1}
   \definecolor{RED}{rgb}{1,0,0}
   \definecolor{GREEN}{rgb}{0,1,0}
   \definecolor{BLUE}{rgb}{0,0,1}
   \definecolor{CYAN}{cmyk}{1,0,0,0}
   \definecolor{MAGENTA}{cmyk}{0,1,0,0}
   \definecolor{YELLOW}{cmyk}{0,0,1,0}
 }

\usepackage{aas_macros}
\usepackage{bbold}

\let\textquotedbl="

\makeatother

\begin{document}

\title{Kinematic Dynamo, Supersymmetry Breaking, and Chaos}

\author{Igor V. Ovchinnikov}

\affiliation{Electrical Engineering Department, University of California at Los
Angeles, Los Angeles, 90095 CA, USA}

\author{Torsten A. Enßlin}

\affiliation{{\small{}Max-Planck-Institut für Astrophysik, Karl-Schwarzschildstr.~1,
85748 Garching, Germany}\\
Ludwig-Maximilians-Universität München, Geschwister-Scholl-Platz{\small{}~}1,
80539 Munich, Germany\\
Technische Universität München, Exzellenzcluster Universe, Boltzmannstr.{\small{}~}2,
85748 Garching, Germany}
\begin{abstract}
The kinematic dynamo (KD) describes the growth of magnetic fields
generated by the flow of a conducting medium in the limit of vanishing
backaction of the fields onto the flow. The KD is therefore an important
model system for understanding astrophysical magnetism. Here, the
mathematical correspondence between the KD and a specific stochastic
differential equation (SDE) viewed from the perspective of the supersymmetric
theory of stochastics (STS) is discussed. The STS is a novel, approximation-free
framework to investigate SDEs. The correspondence reported here permits
insights from the STS to be applied to the theory of KD and vice versa.
It was previously known that the fast KD in the idealistic limit of
no magnetic diffusion requires chaotic flows. The KD-STS correspondence
shows that this is also true for the diffusive KD. From the STS perspective,
the KD possesses a topological supersymmetry and the dynamo effect
can be viewed as its spontaneous breakdown. This supersymmetry breaking
can be regarded as the stochastic generalization of the concept of
dynamical chaos. As this supersymmetry breaking happens in both the
diffusive and the non-diffusive case, the necessity of the underlying
SDE being chaotic is given in either case. The observed exponentially
growing and oscillating KD modes prove physically that dynamical spectra
of the STS evolution operator that break the topological supersymmetry
exist with both, real and complex ground state eigenvalues. Finally,
we comment on the non-existence of dynamos for scalar quantities.
\end{abstract}


\maketitle

\section{Introduction}

The magnetohydrodynamical dynamo has a long scientific history and
many important theoretical insights on it have already been provided
\citep[ and others]{1954RSPTA.247..213B,1955ApJ...122..293P,1964Ge&Ae...4..572B,1970ApJ...162..665P,1980opp..bookR....K,Book_on_Magnetic_Dynamo}.
Roughly speaking, the magnetic dynamo phenomenon is the ability of
a moving conducting medium to generate and/or to sustain a magnetic
field. This phenomenon is wide-spread in astrophysical objects like
galaxies \citep[ for example]{1992ApJ...401..137P,1994A&A...289...94B,1998MNRAS.294..718S,2007A&A...470..539B},
galaxy clusters \citep{1989MNRAS.241....1R,2002A&A...387..383D,2006A&A...453..447E,2014MNRAS.445.3706V},
stars \citep{1975ApJ...198..205P,1984JCoPh..55..461G,1986AN....307...89R,2001ApJ...559.1094C,2002A&A...381..923S,2006A&A...449..451B,2008ApJ...676.1262B},
and planets including Earth \citep{1954RSPTA.247..213B,1955ApJ...122..293P,1963RSPSA.274..274T,1964Ge&Ae...4..572B,1986AN....307...89R,1997Natur.389..371K}.
System-sized ordered fields can be observed in objects with large-scale
ordered flows such as those occurring in rotating galaxies, stars,
and planets, and are usually attributed to the action of the so-called
large-scale, mean-field, or $\alpha-\mbox{\ensuremath{\omega}}$ dynamo.
Systems without a large-scale flow pattern can also harbor dynamos
driven by turbulent fluid motions. These are called small-scale, fluctuating,
or turbulent dynamos and are believed to maintain the magnetic fields
in galaxy clusters. Although the theory of dynamos is relatively mature,
it still provides a field of active and interesting research.

Dynamo theory addresses two different regimes. The first one is the
kinematic dynamo (KD, \citep{1964Ge&Ae...4..572B,1971ApJ...166..627L,1972RSPTA.272..663R,Dynamo_Complex}).
In this regime, the amplified magnetic field is too weak to affect
the flow of the conducting medium. This regime is realized, for example,
in the early stages of galaxy formation. The other regime is the nonlinear
dynamo. Here, the backaction of a sufficiently strongfield on the
flow is not negligible anymore. Such backactions lead eventually to
a saturated dynamo state, with an average stationary magnetic energy
linked to the kinetic energy of the flow. This regime seems to be
realized in the magnetic fields in developed galaxies and galaxy clusters.
This paper, however, focuses solely on the KD limit.

It is known (see, e.g., Ref.~\citep{Motter_Chaos} and references
therein) that the KD in the idealistic diffusionless case requires
chaotic underlying flows of matter (see, e.g., Ref.~\citep{ChaosDynamo}
and references therein). It was not known until this work whether
this relation also holds if magnetic diffusivity is present. Establishing
the chaos-KD relation under diffusivity requires a rigorous mathematical
generalization of the concept of deterministic chaos to stochastic
flows, which was missing so far. The recently found supersymmetric
theory of stochastics (STS) closes this gap \citep{Mine_1,Mine_2}.

The STS is an approximation-free theory of stochastic differential
equations (SDEs). Instead of investigating the ensemble of stochastic
trajectories generated by an SDE, the STS analyzes the actions (or
pullbacks) induced by these trajectories on the elements of the exterior
algebra of the phase space -- the differential forms of various degrees,
in the following called wavefunctions. These wavefunctions can be
averaged over the stochastic noise configurations, as the exterior
algebra, together with the SDE-induced actions on it, is a linear
space in which the stochastic averaging is legitimate. In contrast,
the concept of the stochastically averaged trajectories themselves
can not be straightforwardly defined for non-linear phase spaces.
The stochastically averaged SDE-defined pullback, i.e., the stochastic
evolution operator, provides a complete picture of all aspects of
the stochastic dynamics. For example, the temporal evolution of an
initial $\delta$-functional probability distribution in the phase
space can be explicitly constructed from it.

The KD-STS correspondence that we want to highlight here emerges from
the identification of the fluid flow and the magnetic fields of the
KDs with the flow vector field of an SDE and the (non-supersymmetric)
two-forms (or 1-forms in case one described the magnetic field in
terms of the magnetic vector potential) respectively. This connection
permits the transfer of insights obtained within the STS to the KD
theory and vice versa.

In the STS, all SDEs possess a topological (or De Rahm) supersymmetry.
This means that the stochastic evolution operator commutes\footnote{The infinitesimal stochastic evolution operator is actually ``$\hat{d}$-exact''
which is stricter than a mere commutativity with $\hat{d}$.} with the so-called exterior derivative or De Rahm operator, $\hat{d}$,
of the exterior algebra (see Sec. \ref{sec:N=00003D1-Supersymmetry-of}).
As a mathematical consequence of this (see Sec. \ref{sub:Dynamical-spectra}),
all eigenstates of a time-independent stochastic evolution operator
of a stationary SDE are divided into the finite number (one for each
De Rahm cohomology class) of the zero-eigenvalue (or steady-state)
supersymmetric singlets and the infinite number of the non-supersymmetric
doublets with arbitrary, but bounded from below eigenvalues.

If there are growing eigenstates, the fastest growing eigenstate must
be identified as the ground state of the model. Indeed, if an initial
wavefunction has a contribution by such an eigenstate, this fastest
growing eigenstate will dominate over the initial wavefunction after
a sufficiently long period. The supersymmetry is said to be broken
spontaneously in such a situation because this ground state has non-zero
eigenvalue and thus is non-supersymmetric. 

KDs are known to show growing magnetic fields \citep{ChaosDynamo}.
Thus, the supersymmetry of their corresponding SDEs is spontaneously
broken. In other words, the existence of the growing modes of the
KD is the physical proof that the supersymmetry breaking spectra of
the STS evolution operator are realizable. Moreover, the well established
existence of the oscillating fastest growing KD modes \citep{1987AN....308...89B,1988JFM...197...39R,1220.78120}
demonstrate that the STS spectra with complex ground state eigenvalues
are also realizable. 

The concept of spontaneous supersymmetry breaking in the STS, required
for the existence of the growing KD modes, can actually be identified
with the stochastic generalization of the concept of deterministic
chaos \citep{Mine_1}. For spontaneously broken symmetries, the Goldstone
theorem predicts\footnote{This version of the Goldstone theorem applies to ``spatially extended''
models with infinite-dimensional phase spaces. Fore finite-dimensional
phase spaces, the theorem reduces to the statement of the protected
degeneracy of the ground state, so that there is a ``zero-energy''
excitation that can remember perturbations forever, i.e., the butterfly
effect.} the existence of gapless Goldstone-Nambu excitations, which have
infinite characteristic time-scales. This long-term dynamical behavior
can be regarded as the manifestation of the ubiquitously observed
``chaotic or dynamical long-range order'' that reveals itself in
such phenomena as the butterfly effect \citep{Motter_Chaos}, $\nicefrac{1}{f}$
noise \citep{RevModPhys.53.497}, and the power-law statistics of
various sudden (or instantonic) processes like solar flares \citep{Asc11},
earthquakes \citep{Gut55}, and neuronal avalanches \citep{Beg03}.
Therefore, the long-range phenomena associated with the presence of
the dynamical long-range order must also be present in the KD systems,
which is one of the interesting conclusions that can be drawn from
the KD-STS correspondence discussed in this paper. 

The structure of the paper is as follows: In Sec.~\ref{sec:N=00003D1-Supersymmetry-of},
it is demonstrated that the dynamical (stochastic) equations of the
stationary KD possess topological supersymmetry. In Sec.~\ref{sec:Connection-with-the},
the SDE corresponding to the KD effect is established and the relation
of STS to growing KD modes, chaos, and supersymmetry breaking is elaborated.
Sec.~\ref{sec:Conclusion} concludes the paper.

\section{Supersymmetry of the Kinematic Dynamo Equation\label{sec:N=00003D1-Supersymmetry-of}}

Temporal evolution of the magnetic field within the KD effect is governed
by the induction equation: 
\begin{eqnarray}
\partial_{t}B & = & \partial\times\left(v\times B\right)+\eta\triangle B.\label{StochInductionEq}
\end{eqnarray}
Here, $\partial_{t}\equiv\partial/\partial t$, $\partial=\{\partial_{i}\equiv\partial/\partial x^{i},x^{i}=x,y,z,i=1,2,3$\},
is the gradient operator, $\triangle=\partial^{2}=\partial_{i}\partial_{i}$,\footnote{The summation over repeated indices is assumed throughout the paper.}
is the Laplace operator, $\times$ denotes the vector product of two
vectors so that $\partial\times$ is the curl of a vector, $B=\{B^{i},i=1,2,3\}$
is the magnetic field, $v=\{v^{i},i=1,2,3\}$ is the vector field
of the underlying flow velocity of the conducting medium, and $\eta=1/(\sigma\mu)$
is the magnetic diffusivity with $\sigma$ and $\mu$ being the electrical
conductivity and permeability. The first term in the r.h.s of Eq.~\eqref{StochInductionEq}
represents the well-known magnetohydrodynamical phenomenon of magnetic
fields being frozen into the conducting medium, whereas the second
term describes the magnetic field diffusion.

Our first goal is to translate Eq.~\eqref{StochInductionEq} into
the coordinate-free language of exterior algebra used by the STS.
Instead of the vector $B$, we use differential forms of second degree
to describe magnetic fields. Such 2-forms provide the coordinate-free
representation of the same object, 
\begin{eqnarray}
F & = & \frac{1}{2!}F_{ij}dx^{i}\wedge dx^{j}=\hat{d}A.\label{Vector Potential}
\end{eqnarray}
Here, $A=A_{i}dx^{i}$ is the 1-form of the magnetic vector potential,
$\hat{d}=dx^{i}\wedge\partial/\partial x^{i}$ is the exterior derivative
or the De Rahm operator, and $\wedge$ is the wedge or antisymmetric
product of differentials. In components, the antisymmetric contra-variant
tensor, $F_{ij}$, called the magnetic field tensor, is given as:
\begin{eqnarray*}
F_{ij} & = & \partial_{i}A_{j}-\partial_{j}A_{i}=\epsilon_{ijk}B^{k}=\left(\begin{array}{ccc}
0 & B^{z} & -B^{y}\\
-B^{z} & 0 & B^{x}\\
B^{y} & -B^{x} & 0
\end{array}\right),
\end{eqnarray*}
where $\epsilon_{ijk}$ is the antisymmetric Levi-Cevita tensor.

Eq.~\eqref{StochInductionEq} in components is

\[
\partial_{t}B^{i}=\epsilon^{ipq}\partial_{p}\epsilon_{qkl}v^{k}B^{l}+\eta\triangle B^{i}.
\]
We work in the Eulcidian metric. Thus, lowering and raising the indices
has no effect on the values of the components of the antisymmetric
tensor, e.g., $\epsilon_{ijk}=\epsilon^{ijk}$. With the use of the
following identity,

\begin{equation}
\epsilon_{qkl}\epsilon^{ipq}=\mathrm{det}\left(\begin{array}{cc}
\delta_{k}^{i} & \delta_{k}^{p}\\
\delta_{l}^{i} & \delta_{l}^{p}
\end{array}\right),\label{eq:determ_2}
\end{equation}
where $\delta_{j}^{i}$ is the Kronecker-delta and with $\partial_{i}B^{i}=0,$
Eq. (\ref{StochInductionEq}) can be rewritten as:

\[
\partial_{t}B^{i}=-\left(\partial_{j}v^{j}\right)B^{i}+B^{j}v_{'j}^{i}+\eta\triangle B^{i},
\]
where $v_{'j}^{i}=\partial_{j}v^{i}$. Using now

\[
B^{i}=\frac{1}{2}\epsilon^{ikl}F_{kl},
\]
the induction equation can be further rewritten as:

\begin{eqnarray*}
\partial_{t}\frac{1}{2}\epsilon^{ikl}F_{kl} & = & -\partial_{j}v^{j}\frac{1}{2}\epsilon^{ikl}F_{kl}+\frac{1}{2}\epsilon^{jkl}F_{kl}v_{'j}^{i}\\
 &  & +\eta\triangle\frac{1}{2}\epsilon^{ikl}F_{kl}.
\end{eqnarray*}
Multiplying both sides of this Eq.~by $\epsilon_{iab}$ and summing
over index $i$, we get:

\[
\partial_{t}F_{ab}=-\partial_{j}v^{j}F_{ab}+\frac{1}{2}\epsilon_{iab}\epsilon^{jkl}F_{kl}v_{'j}^{i}+\eta\triangle F_{ab}.
\]
With the help of the identity

\[
\epsilon_{iab}\epsilon^{jkl}=\mathrm{det}\left(\begin{array}{ccc}
\delta_{i}^{j} & \delta_{a}^{j} & \delta_{b}^{j}\\
\delta_{i}^{k} & \delta_{a}^{k} & \delta_{b}^{k}\\
\delta_{i}^{l} & \delta_{a}^{l} & \delta_{b}^{l}
\end{array}\right),
\]
one arrives at

\begin{eqnarray*}
\partial_{t}F_{ab} & = & -\partial_{j}v^{j}F_{ab}+\frac{1}{2}\left(2F_{ab}v_{'j}^{j}-2F_{jb}v_{'a}^{j}-2F_{aj}v_{'b}^{j}\right)\\
 &  & +\eta\triangle F_{ab},
\end{eqnarray*}
or

\[
\partial_{t}F_{ab}=-\left(v^{j}\partial_{j}F_{ab}+v_{'a}^{j}F_{jb}+F_{aj}v_{'b}^{j}\right)+\eta\triangle F_{ab}.
\]
Turning now to the coordinatee-free object $F$ in Eq.~\eqref{Vector Potential},
Eq.~\eqref{StochInductionEq} takes the form
\begin{eqnarray}
\partial_{t}F & = & -\hat{H}_{\mathrm{KD}}F,\:\;\hat{H}_{\mathrm{KD}}=\hat{\mathcal{L}}_{v}-\eta\hat{\triangle},\label{EqNew1}
\end{eqnarray}
where the Lie derivative along $v$ is defined as 
\begin{eqnarray*}
\hat{\mathcal{L}}_{v} & = & v^{i}\partial_{i}+v_{'j}^{i}dx^{j}\wedge\hat{\imath}_{i}.
\end{eqnarray*}
Here, $\hat{\imath}_{i}$ is the interior multiplication acting on
a differential form $\psi=\frac{1}{k!}\psi_{i_{1}...i_{k}}dx^{i_{1}}\wedge...\wedge dx^{i_{k}}$
as 
\begin{eqnarray*}
\hat{\imath}_{i}\psi & = & \frac{1}{(k-1)!}\psi_{ii_{2}...i_{k}}dx^{i_{2}}\wedge...\wedge dx^{i_{k}}.
\end{eqnarray*}
The resulting Eq.~\eqref{EqNew1} is rather natural. As we already
mentioned, the first term in the r.h.s.~of Eq.~\eqref{StochInductionEq}
describes the infinitesimal temporal evolution of the magnetic field
``frozen'' into the conducting medium. This freezing of the magnetic
field is the well-known magnetohydrodynamical effect. In the coordinate-free
setting, the frozen field corresponds to the evolution solely due
to the flow along $v$ and this evolution is given by the Lie derivative,
which is also known as the physical derivative.

The next step towards STS is to establish the supersymmetric structure
of the KD evolution operator, $\hat{H}_{\mathrm{KD}}$. 

First, we recall the Cartan formula, 
\begin{eqnarray}
\hat{\mathcal{L}}_{v} & = & [\hat{d},\hat{\imath}_{v}],\label{CartanForm}
\end{eqnarray}
where $\hat{\imath}_{v}=v^{i}\,\hat{\imath}_{i}$ the interior multiplication
with $v$ and the square brackets denote the bi-graded commutator.
This is defined by 
\begin{eqnarray*}
[\hat{X},\hat{Y}] & = & \hat{X}\hat{Y}-(-1)^{\mathrm{deg}\hat{X}\mathrm{\cdot deg}\hat{Y}}\hat{Y}\hat{X},
\end{eqnarray*}
where $\mathrm{deg}$ is the degree of the operator, i.e., the number
of $dx$'s minus the number of $\hat{\imath}$'s. For example, $\mathrm{deg}\,\hat{d}=1$
and $\mathrm{deg}\,\hat{\imath}_{v}=-1$ so that the bi-graded commutator
in Eq.~\eqref{CartanForm} is actually an anti-commutator. 

Second, the Laplace operator can be given by 
\begin{eqnarray}
\hat{\triangle} & = & -[\hat{d},\hat{d}^{\dagger}],\label{Lapl}
\end{eqnarray}
where $\hat{d}^{\dagger}=-\hat{\imath}_{i}\delta^{ij}\partial_{j}$
is the Hodge conjugate of the exterior derivative with respect to
the Euclidian metric.

Substituting Eqs.~\eqref{Lapl} and \eqref{CartanForm} into Eq.~\eqref{EqNew1},
one finds the explicitly supersymmetric form of the KD evolution operator,
\begin{eqnarray}
\hat{H}_{\mathrm{KD}} & = & [\hat{d},\hat{\bar{d}}],\label{SUSYHOp}
\end{eqnarray}
where 
\begin{eqnarray*}
\hat{\bar{d}} & = & \hat{\imath}_{v}+\eta\hat{d}^{\dagger}.
\end{eqnarray*}
The notational similarity of the exterior derivative $\hat{d}$ and
the evolution-defining $\hat{\bar{d}}$ is motivated by the fact that
for a purely diffusive dynamics, with $\hat{H}=-\eta\hat{\triangle}$,
these operators are the Hodge conjugates to each other (up to a diffusion
constant) with respect to the Euclidian metric, $\hat{\bar{d}}=\eta\hat{d}^{\dagger}$.

We now show the supersymmetry of $\hat{H}_{\mathrm{KD}}$. We recall
that due to the nil-potency of the exterior derivative, $\hat{d}^{2}=0$,
one has

\begin{flushleft}
\begin{eqnarray}
[\hat{d},[\hat{d},\hat{X}]] & = & 0,\label{Nilpotency operator}
\end{eqnarray}
for any $\hat{X}$. Thus, an immediate consequence of Eq.~\eqref{Nilpotency operator}
and the ``$\hat{d}$-exactness'' of the KD evolution operator in
Eq.~\eqref{SUSYHOp}, is that it commutes with $\hat{d}$:
\par\end{flushleft}

\begin{eqnarray}
[\hat{d},\hat{H}_{\mathrm{KD}}] & = & 0.\label{eq:CommutativityWithD}
\end{eqnarray}
This suggests that $\hat{d}$ is a symmetry of the KD dynamics. Since
$\hat{d}=dx^{i}\wedge\partial/\partial x^{i}$ removes a bosonic (commuting)
variable (an $x^{i}$ is removed by $\partial/\partial x^{i}$) and
replaces it by a fermionic one (an anti-commuting $dx^{i}\wedge$
is added) it converts bosonic variables into fermionic ones. Therefore,
the symmetry of the evolution operator $\hat{H}_{\mathrm{KD}}$ with
respect to $\hat{d}$ can be regarded as a supersymmetry since an
exchange of a boson by a fermion in the wavefunction does not change
its dynamics.

As a consequence of its supersymmetry conserving dynamics, the KD
is described equivalently in terms of the magnetic field tensor and
in terms of the vector potential in Eq. (\ref{Vector Potential}),

\begin{eqnarray}
\partial_{t}A & = & -\hat{H}_{\mathrm{KD}}A\nonumber \\
\Rightarrow\partial_{t}\hat{d}A & = & -\hat{d}\hat{H}_{\mathrm{KD}}A\nonumber \\
\Rightarrow\partial_{t}\hat{d}A & = & -\hat{H}_{\mathrm{KD}}\hat{d}A\nonumber \\
\Rightarrow\partial_{t}F & = & -\hat{H}_{\mathrm{KD}}F.\label{eq:AandFdynamics}
\end{eqnarray}
From a traditional KD perspective, it might come as a surprise that
the magnetic field tensor and magnetic vector potential obey exactly
the same evolution equation, despite being mathematically different,
although related objects. From a STS perspective, this equivalence
just shows the consistence of our calculations. In fact, the descriptions
of the magnetic field evolution in terms of $A$ and $F$ must be
equivalent. This suggests on its own that the exterior derivative
(connecting $A$ and $F=\hat{d}A$) must be commutative with the evolution
operator. Thus, we could just as well have guessed the existence of
this supersymmetry from the very outset. 

Note, however, that Eq.~\eqref{SUSYHOp} not only implies the commutativity
of the evolution operator with the exterior derivative, Eq.~\eqref{eq:CommutativityWithD},
it also implies that all the supersymmetric eigenstates have zero
eigenvalue, as we discuss in the next section.

The close relation between supersymmetry, invariance under the action
of $\hat{d}$, and algebraic topology, relations of topological sets
with their boundaries, which we mentioned in the introduction, was
first established in Ref.~\citep{Witten_SUSY}. This relation had
resulted in the discovery of Witten-type topological or cohomological
field theories (see, e.g., Refs.~\citep{TFT_Frankel,TFT_Labastida,TFT_Witten_Sigma,TFT_Witten}
and references therein, as well as Ref.~\citep{TFT_Review} for a
review).\footnote{In the path integral representation of cohomological field theories,
the $\hat{d}$-supersymmetry is denoted as $Q$ and is identified
as the topological supersymmetry or as the gauge-fixing Becchi-Rouet-Stora-Tyutin
symmetry.} The STS that we are going to discuss next can be looked at as a member
of this class of theories.\footnote{In a full-fledged cohomological field theory, one is interested only
in the supersymmetric ground states of the model. In the STS, on the
other hand, one is primarily interested in the ground states, which
for chaotic systems are non-supersymmetric as we will argue. From
this point of view, the STS can be recognized as a cohomological theory
only in a generalized sense.}

\section{Connection between the KD theory and the STS\label{sec:Connection-with-the}}

\subsection{Corresponding stochastic system }

In this section, we will show that the stationary KD equation including
magnetic diffusivity is the STS stochastic evolution operator of a
certain SDE. To find this corresponding SDE, let us address here a
more general SDE, establish its stochastic evolution operators, and
compare it with $\hat{H}_{\mathrm{KD}}$. This general SDE can be
thought of as one describing the trajectory of a test particle propagating
along the fluid flow while its velocity is also subject to Gaussian
white noise, 
\begin{eqnarray}
\dot{x}^{i}(t) & = & v^{i}(x(t))+(2\Theta)^{1/2}e_{a}^{i}(x(t))\xi^{a}(t).\label{Corresponding SDE}
\end{eqnarray}
Here, $x$ is the particle (or phase-space) position, $v=\{v^{i}\}$
is the stationary part of the flow vector field, $e_{a}(x)=\{e_{a}^{i}(x);i,a=1,2,3\}$
is a set of vector fields that can be characterized as the ``prime
vectors'' of flow fluctuations at x and that are assumed position-dependent
for now, $\xi^{a}(t)\in\mathbb{R}^{1}$ with $a\in\left\{ 1,2,3\right\} $
are the Gaussian white fluctuations with the standard stochastic averages,
\begin{eqnarray}
\langle\xi^{a}(t)\rangle_{\text{noise}} & = & 0\text{ and }\label{NoiseAverages}\\
\langle\xi^{a}(t)\xi^{b}(t')\rangle_{\mathrm{noise}} & = & \delta^{ab}\delta(t-t'),\label{eq:NoiseAverages1}
\end{eqnarray}
and $\Theta$ is the intensity or temperature of these fluctuations.
Getting a bit ahead, the flow vector field of the SDE corresponding
to the KD equation will turn out to be the same as in the original
KD problem. This is the reason why we do not introduce a new notation
for the flow in Eq.\ref{Corresponding SDE}.

First, an evolution operator $M_{t't}^{*}$ is defined, which describes
how the dynamics of the system acts on elements of the exterior algebra
of the position (or phase) space. The elements of the exterior algebra
form a Hilbert space, the space of differential forms of all degrees,
$\varOmega(X)$. We recall that our magnetic field tensor and magnetic
vector potential are among such elements, they are 2- and 1-forms
respectively. Mathematically, $M_{t't}^{*}$ is the pullback of the
inverse, finite-time evolution diffeomorphism. For a fixed noise configuration,
this evolution operator is

\begin{equation}
M_{t't}^{*}=\mathcal{T}e^{-\int_{t'}^{t}d\tau\mathcal{\hat{L}}_{v(\tau)}},\label{eq:Pullback}
\end{equation}
where $\mathcal{T}$ expresses chronological ordering of the terms
to its right (see below). The time-dependent flow vector field is
the entire r.h.s. of Eq.~\eqref{Corresponding SDE}, 

\[
v(t)=v^{i}+(2\Theta)^{1/2}e_{a}^{i}\xi^{a}(t),
\]
and the corresponding time dependent Lie derivative of this flow is

\begin{equation}
\mathcal{\hat{L}}_{v(t)}=\mathcal{\hat{L}}_{v}+(2\Theta)^{1/2}\xi^{a}(t)\mathcal{\hat{L}}_{e_{a}}.\label{eq:LinearityOfLieDerivative}
\end{equation}
Since the Lie derivative is linear in its argument, a stochastic term
is just added to the time independent Lie derivative given by Eq.~\eqref{CartanForm}.
The operator of the chronological ordering $\mathcal{T}$ is needed
in Eq.~\eqref{eq:Pullback} because in general instances of $\mathcal{\hat{L}}_{v(\tau)}$
at different times do not commute with each other. Eq.~\eqref{eq:Pullback}
follows immediately from one of the definitions of the Lie derivative,
which is the infinitesimal pullback along a vector field. 

Secondly, the finite-time evolution operator $M_{t't}^{*}$ is averaged
over all the configurations of the stochastic noise:

\begin{equation}
\hat{M}_{tt'}=\left\langle M_{t't}^{*}\right\rangle _{\text{noise}}.\label{eq:DefinitionStochPullback}
\end{equation}
As $M_{t't}^{*}$ is a linear operator acting on $\varOmega(X)$,
the Hilbert space of all differential forms in which superpositions
are possible, $\hat{M}_{tt'}$ represents a mathematically meaningful
average. This averaged finite-time evolution operator describes how
wavefunctions evolve in time under the action of a dynamical system:

\begin{equation}
\psi(t)=\hat{M}_{tt'}\psi(t'),\,\psi\in\varOmega(X).
\end{equation}

The infinitesimal stochastic evolution of the wave function can be
given via the stochastic evolution equation 

\begin{equation}
\partial_{t}\psi(t)=-\hat{H}\psi(t),\label{eq:FPeq}
\end{equation}
where the stochastic evolution operator (SEO) is defined by 
\begin{eqnarray}
\hat{H} & = & \lim_{\delta t\to0}\frac{\hat{1}_{\Omega}-\hat{M}_{(t+\delta t)t}}{\delta t}.\label{eq:Infinitesemal_H}
\end{eqnarray}

STS investigates the evolution of the complete set of forms $\psi\in\Omega(X)$,
whereas in the KD theory usually only the 2-forms $F\in\Omega^{2}(X)\subset\Omega(X)$
are of interest (or 1-forms $A\in\Omega^{1}(X)\subset\Omega(X)$ if
one chooses to work with the vector potential).\footnote{In the STS, the wavefunctions have the meaning of generalized probability
distributions in the coordinate-free setting, e.g, the top differential
forms can be viewed as the total probability distributions (when strictly
positive and normalized to one) whereas the lower-degree differential
forms can be looked upon as the conditional probability distributions
(under the same conditions). } For example, the total probability density 3-form $\psi(t)=\rho(x,t)\,dx^{1}\wedge dx^{2}\wedge dx^{3}\in\varOmega^{3}(X)$
to find the system in a specific phase space volume around location
$x$ at time $t$ evolves according to Eq.~\eqref{eq:FPeq} restricted
to $\varOmega^{3}(X)$. The correspondingly restricted SEO is the
conventional Fokker-Planck operator and will here be denoted as $\hat{H}^{(3)}:\varOmega^{3}(X)\rightarrow\varOmega^{3}(X)$.

We now work out the SEO of our SDE \eqref{Corresponding SDE}. Using
Eqs.~\eqref{eq:Pullback}, \eqref{eq:LinearityOfLieDerivative},
\eqref{eq:DefinitionStochPullback}, and \eqref{eq:Infinitesemal_H},
the formal definition of the chronologically ordered exponentiation
of the operator in Eq.~\eqref{eq:Pullback}, 
\begin{eqnarray}
\mathcal{T}e^{-\int_{t'}^{t}d\tau\mathcal{\hat{L}}_{v(\tau)}} & = & \hat{1}-\int_{t'}^{t}d\tau\mathcal{\hat{L}}_{v(\tau)}\label{ChonTaylor}\\
 &  & +\int_{t'}^{t}d\tau_{1}\mathcal{\hat{L}}_{v(\tau_{1})}\int_{t'}^{\tau_{1}}d\tau_{2}\mathcal{\hat{L}}_{v(\tau_{2})}-...,\nonumber 
\end{eqnarray}
and the stochastic averages of the Gaussian white noise from Eqs.~\eqref{NoiseAverages}
and \eqref{eq:NoiseAverages1}, one readily finds\footnote{Note that on derivation of Eq.~\eqref{HOp}, the last term acquires
the factor 
\begin{eqnarray*}
\lim_{\delta t\to0}2\delta t^{-1}\int_{t}^{t+\delta t}d\tau_{1}\int_{t}^{\tau_{1}}d\tau_{2}\delta(\tau_{1}-\tau_{2})
\end{eqnarray*}

\begin{eqnarray*}
 & = & \lim_{\delta t\to0}\delta t^{-1}\int_{t}^{t+\delta t}d\tau_{1}=1.
\end{eqnarray*}
The first equality here follows from $\int_{t}^{\tau_{1}}d\tau_{2}\delta(\tau_{1}-\tau_{2})=\nicefrac{1}{2}$,
which is a consequence of defining the $\delta$-distribution to be
the limit of a narrowing sequence of symmetric functions.} 
\begin{eqnarray}
\hat{H} & = & \hat{\mathcal{L}}_{v}-\Theta\hat{\mathcal{L}}_{e_{a}}\hat{\mathcal{L}}_{e_{a}}.\label{HOp}
\end{eqnarray}

In order to establish the supersymmetric structure of the SEO, we
recall that the bi-graded commutator with the exterior derivative
is a bi-graded differentiation, 
\begin{equation}
[\hat{d},\hat{X}\hat{Y}]=[\hat{d},\hat{X}]\hat{Y}+(-1)^{\mathrm{deg}\hat{X}}\hat{X}[\hat{d},\hat{Y}].\label{Differention_by_D}
\end{equation}

Thus, the bi-graded commutator with $\hat{d}$ and the Cartan formula
\eqref{CartanForm} imply that 
\begin{eqnarray}
\hat{\mathcal{L}}_{e_{a}}\hat{\mathcal{L}}_{e_{a}} & = & [\hat{d},\imath_{e_{a}}\hat{\mathcal{L}}_{e_{a}}].\label{LieDouble}
\end{eqnarray}
Thus

\begin{eqnarray}
\hat{H} & = & [\hat{d},\hat{\bar{d}}],\qquad\hat{\bar{d}}=\hat{\imath}_{v}-\Theta\imath_{e_{a}}\hat{\mathcal{L}}_{e_{a}}.
\end{eqnarray}
The SEO $\hat{H}$ of our SDE \eqref{Corresponding SDE} is therefore
the KD evolution operator $\hat{H}_{\mathrm{KD}}$ from the previous
section,
\begin{eqnarray}
\hat{H}= & \hat{H}_{\mathrm{KD}} & ,\text{ for }e_{a}^{i}(x)=\delta_{a}^{i}\text{ and }\Theta=\eta.
\end{eqnarray}
In other words, the KD evolution operator is the SEO of the following
SDE,

\begin{eqnarray}
\dot{x}^{i} & = & v^{i}(x)+(2\eta)^{1/2}\xi^{i}(t),\label{eq:CorrespondingSDEREd}
\end{eqnarray}
where $\xi^{i}(t)=\delta_{a}^{i}\xi^{a}(t)$ is additive Gaussian
white noise.

\subsection{Dynamical spectra in STS\label{sub:Dynamical-spectra}}

STS investigates the eigenspectrum of wavefunctions of the exterior
algebra under the action of a SDE. As we have seen, the KD modes are
a subset of these wavefunctions for a suitably constructed SDE and
therefore the STS classification of dynamical systems directly applies
to KDs. 

The stochastic evolution equation \eqref{eq:FPeq} is linear. Thus,
the time evolution of any wavefunction, $\psi$, can be constructed
from the complete eigensystem of the SEO $\hat{H}$, 
\begin{align}
\psi(t) & =\sum_{\alpha}a_{\alpha}\,\alpha\,e^{-\mathcal{E_{\alpha}}t}\mbox{, with}\nonumber \\
\hat{H}\alpha & =\mathcal{E_{\alpha}}\alpha\mbox{ and}\nonumber \\
a_{\alpha} & =\langle\alpha|\psi\rangle=\int_{X}\psi\wedge\bar{\alpha}.\label{eq:LinearEvolution}
\end{align}
The set of eigenvectors $\left\{ \alpha\right\} $ of $\hat{H}$ form
a complete bi-orthogonal basis (see below), such that $\hat{1}_{\Omega}=\sum_{\alpha}|\alpha\rangle\langle\alpha|$
permits the decomposition of any wavefunction in $\Omega(X)$ into
the eigenmodes. The eigenvalues determine whether eigenmodes are growing
($\mathrm{Re}\,\mathcal{E}_{\alpha}<0$), decaying ($\mathrm{Re}\,\mathcal{E}_{\alpha}>0$),
oscillating ($\mathrm{Im}\,\mathcal{E}_{\alpha}\neq0$), or stationary
($\mathcal{E}_{\alpha}=0$). As the evolution of any wavefunction
is fully determined by the behavior of the eigenmodes it is composed
of, it is sufficient to study the eigensystem of the SEO to understand
the properties of the corresponding dynamical system. 

Analogously, KD theory concentrates on the spectrum of dynamo eigenmodes,
which are just a subset of $\left\{ \alpha\right\} $.

The eigensystem of the SEO has the following properties:\footnote{We discuss the properties of the SEO under the assumption of a compact
phase space. This assumption is made for simplicity in order to avoid
complications of purely mathematical origin that appear in the non-compact
setting, in which the eigensystem will depend on the choice of the
class of functions that we believe constitute the Hilbert space. From
the physical point of view, this assumption is not a restriction.
For example, a classical model related to the KD phenomenon is the
ABC flow defined on a 3-torus and planetary, stellar, and galactic
dynamos act within finite volumes without flows reaching infinity.} First of all, for non-zero temperatures, $\Theta\geq0$, the SEO
is elliptic, thus (the real part of) its spectrum is bounded from
below.\footnote{In fact, this must be true even for zero temperatures as follows from
the analysis of the spectra of transfer operators in the dynamical
systems theory.} Secondly, the SEO is real, hence its eigenvalues are either real
or come in complex conjugate pairs known in the dynamical systems
theory as the Ruelle-Pollicott resonances. This property of its spectra
implies that the SEO is pseudo-Hermitian \citep{pseudo_Hermitian}.
As a pseudo-Hermitian operator, the SEO's eigensystem is complete
and bi-orthogonal:

\begin{eqnarray*}
\hat{H}\alpha & = & \mathcal{E_{\alpha}}\alpha\mbox{ and }\bar{\alpha}\hat{H}=\bar{\alpha}\mathcal{E_{\alpha}}\mbox{, with}\\
\langle\alpha|\beta\rangle & = & \int_{X}\beta\wedge\bar{\alpha}=\delta_{\alpha\beta}.
\end{eqnarray*}
Here, the kets $|\alpha\rangle\equiv\alpha\in\Omega=\Omega(X)$ and
bras $\langle\alpha|\equiv\bar{\alpha}=\sum_{\beta}\eta_{\beta\alpha}\star(\beta^{*})\in\Omega$,
with $\star:\Omega^{k}\rightarrow\Omega^{D-k}$ being the Hodge star
operator, are the differential forms of the corresponding right and
left eigenfunctions of the SEO. The bras and kets are related through
the non-trivial Hilbert space metric $\eta_{\beta\alpha}$, which
is the inverse of the overlap matrix of the kets, $\sum_{\beta}\eta_{\beta\alpha}\int_{x}\gamma\wedge\star(\beta^{*})=1_{\alpha\gamma}$. 

The operator of the degree of a differential form $\hat{k}=dx^{i}\wedge\hat{\imath}_{i}$,
with $\hat{k}\psi=k\psi$ and $\psi\in\Omega^{k}(X)$, commutes with
$\hat{H}$,\footnote{This is the trivial consequence of the fact that the degree of the
stochastic evolution operator is zero, $\mathrm{deg}\hat{H}=0$.} thus the degree of an eigenstate of $\hat{H}$ is a good quantum
number, $\hat{k}|\alpha\rangle=k_{\alpha}|\alpha\rangle$. Or in other
words, the dynamics specified by $\hat{H}$ do not convert between
different $k$-forms, but evolve separately within each $\Omega^{k}$-subspace
of $\Omega$. We denote as $\hat{H}^{(k)}$ the projection of $\hat{H}$
on $\Omega^{k}$, so that the block diagonal structure of the SEO
can be expressed as $\hat{H}=\mathrm{diag}(\hat{H}^{(D)},...,\hat{H}^{(0)})$.

Corresponding bras and kets have complimentary degrees: if $|\alpha\rangle\equiv\alpha\in\Omega^{k}$
then $\langle\alpha|\equiv\bar{\alpha}\in\Omega^{D-k}$ if $D$ is
the dimension of the phase space. Otherwise the norm of this eigenstate,
$\langle\alpha|\alpha\rangle$, would vanish. Much like in quantum
theory, the bra-ket combination $P_{\alpha}=\bar{\alpha}\wedge\alpha\in\Omega^{D}$
has the meaning of the total probability distribution associated with
this eigenstate and its norm $\int_{X}P_{\alpha}=\int_{X}\alpha\wedge\bar{\alpha}=\langle\alpha|\alpha\rangle$
should be strictly positive for non-trivial states. Consequently,
$\langle\alpha|\equiv\bar{\alpha}\in\Omega^{D-k}$.

\begin{figure*}[t]
\includegraphics[width=0.5\textwidth]{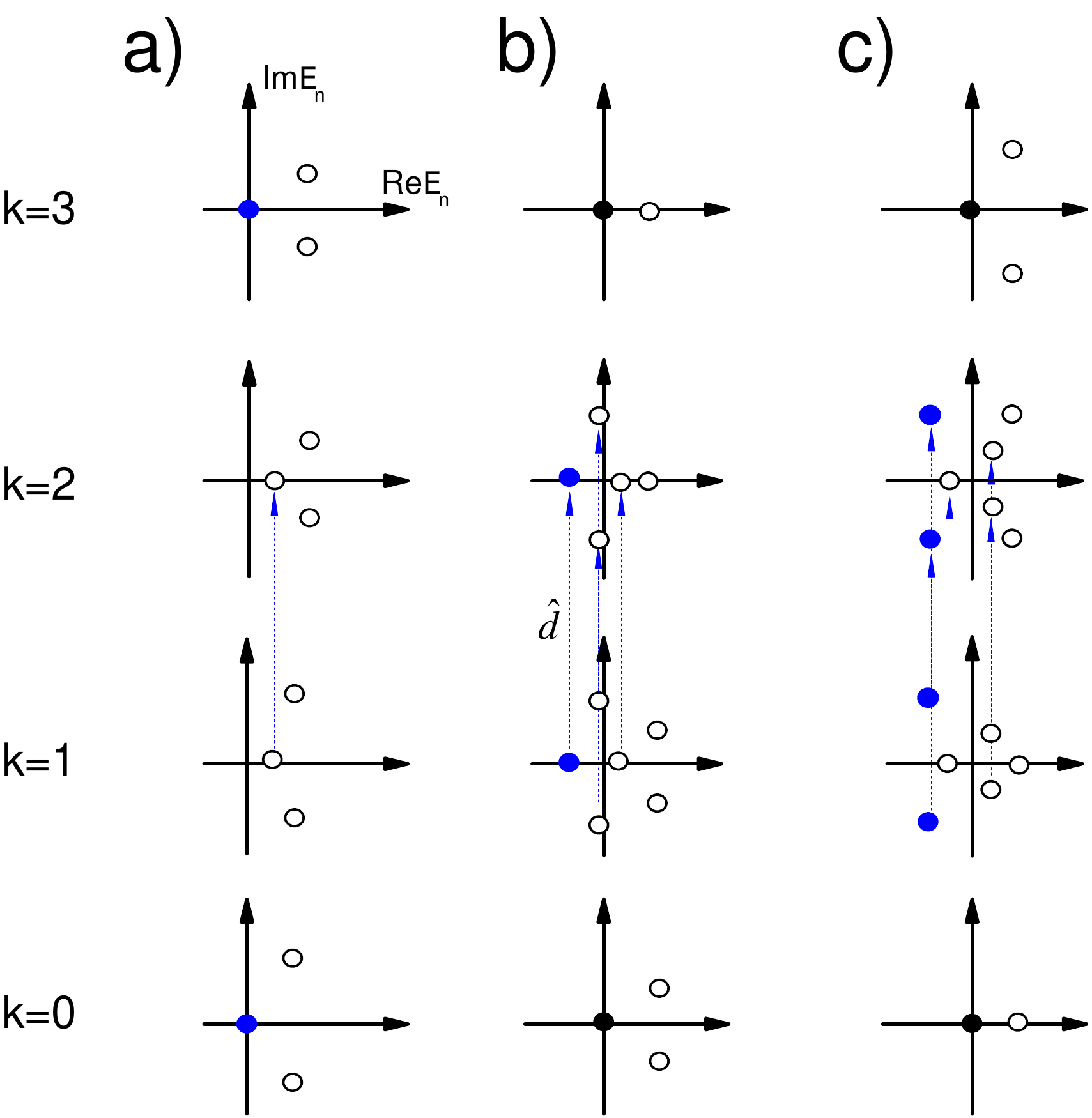}\textcolor{black}{\caption{\label{fig:The-three-possible}(color online) The three possible spectra
of the stochastic evolution operator of the STS of a model with the
phase space being a three-dimentional compact sphere. Such models
have only two supersymmetric eigenstates (big dots at the origin)
in one-to-one correspondence with the two De Rahm cohomology classes
of the 3D sphere in the zeroth (k=0) and the third (k=3) degree. Spectrum
(a) corresponds to the unbroken topological supersymmetry because
both ground states are supersymmetric with one of them (for $k=3$)
being the state of thermodynamic equilibrium. The other two spectra
(b and c) correspond to the ``chaotic'' dynamics, i.e., to the situations
when the ground states (big leftmost dots) are non-supersymmetric
because they have non-zero eigenvalues. \textcolor{black}{The eigenstates
of the stochastic evolution operator that represent the modes of the
magnetic field in the KD theory are the non-$\hat{d}$-symmetric pairs
of 1- and 2-forms connected by the dashed lines representing the action
of the d-operator. Note that the SEO spectra for 1- and 2-forms may
also have other eigenstates (with positive real parts of their eigenvalues)
that can not be associated with the magnetic field. Those from $\Omega^{0}$
can be though of as the eigenstates of the gauge potential $\varphi\in\Omega^{0}$
with which one can perform the gauge-invariance transformation of
the vector potential, $A\to A+\hat{d}\varphi$, without changing the
magnetic field. The additional non-$\hat{d}$-closed eigenmodes in
$\Omega^{2}$ are unphysical because they correspond to the magnetic
field configurations such that $\hat{d}F=\left(\mathrm{div}B\right)d^{3}x\protect\ne0$,
i.e., they contain magnetic monopoles.}}
}
\end{figure*}

The supersymmetric structure of the SEO separates all eigenstates
into two groups: 
\begin{itemize}
\item Almost all eigenstates come in non-supersymmetric ``bosonic-fermionic''
pairs, $|\alpha\rangle$ and $|\alpha'\rangle=\hat{d}|\alpha\rangle$,\footnote{The attempt to continue this construction of eigenstates of the form
$\hat{d}^{n}|\alpha\rangle$ terminates because of the nilpotency
of the exterior derivative, $\hat{d}^{2}=0$.} that have the same eigenvalue because $\hat{H}$ commutes with $\hat{d}$.\footnote{In the KD context, the vector potential $A$ and the magnetic field
tensor $F=\hat{d}\,A$ are an example of the bosonic-fermionic relation
between pairs of non-supersymmetric eigenstates.} 
\item Some of the eigenstates are the supersymmetric singlets $|\theta\rangle$
\textcolor{black}{that are non-trivial in the De Rahm cohomology.
Supersymmetry of a state $|\theta\rangle$ means that it is $\hat{d}$-closed,
$\hat{d}|\theta\rangle=0$, but not $\hat{d}$-exact, meaning that
no $\theta'$ exists such that $|\theta\rangle=\hat{d}|\theta'$$\rangle$}.

\end{itemize}
The $\hat{d}$-exact form of the SEO, i.e., $\hat{H}=[\hat{d},\hat{\bar{d}}]$,
implies that all eigenstates with non-zero eigenvalue are non-supersymmetric.
Indeed, for $\hat{H}|\alpha\rangle=\mathcal{E_{\alpha}}|\alpha\rangle$
and $\mathcal{E_{\alpha}}\ne0$ we have two possibilities:
\begin{enumerate}
\item The first possibility is that $\hat{d}|\alpha\rangle\ne0$. In this
case the statement is trivial because $\hat{H}$ commutes with $\hat{d}$
and consequently the state $|\alpha'\rangle=\hat{d}|\alpha\rangle\ne0$
has the same eigenvalue, $\hat{H}|\alpha'\rangle=\mathcal{E_{\alpha}}|\alpha'\rangle$,
and we have a non-supersymmetric ``boson-fermion'' pair, $|\alpha\rangle$
and $|\alpha'\rangle$. 
\item \textcolor{black}{The other possibility is that $\hat{d}|\alpha\rangle=0$.
Then, using $\hat{H}=\hat{\bar{d}}\hat{d}+\hat{d}\hat{\bar{d}}$ and
the fact that $\hat{d}|\alpha\rangle=0$ we have $\mathcal{E}_{\alpha}|\alpha\rangle=(\hat{\bar{d}}\hat{d}+\hat{d}\hat{\bar{d}})|\alpha\rangle=\hat{d}\hat{\bar{d}}|\alpha\rangle$,
thus $|\alpha\rangle=\hat{d}|\tilde{\alpha}\rangle,$ with $|\tilde{\alpha}\rangle=\hat{\bar{d}}|\alpha\rangle/\mathcal{E}_{\alpha}$.
Up to a ``gauge'', i.e., up to a $\hat{d}$-closed piece, $|\tilde{\alpha}\rangle=|\tilde{\alpha}'\rangle+|\tilde{\alpha}''\rangle$
with $\hat{d}|\tilde{\alpha}''\rangle=0$ is an eigenstate of the
SEO with the same eigenvalue as $|\alpha\rangle$. }\footnote{\textcolor{black}{To see this, one resolves $|\tilde{\alpha}\rangle$
in the eigenstates of $\hat{H}^{(k-1)}$, with k being the degree
of $|\alpha\rangle$, $|\tilde{\alpha}\rangle=\sum_{\tilde{\alpha}_{1}}C_{\tilde{\alpha}_{1}}|\tilde{\alpha}_{1}\rangle+\sum_{\tilde{\alpha}_{2}}C_{\tilde{\alpha}_{2}}|\tilde{\alpha}_{2}\rangle$,
where the labels $\tilde{\alpha}_{1}$ and $\tilde{\alpha}_{2}$ run
over the eigenstates such that $\hat{d}|\tilde{\alpha}_{1}\rangle\ne0$
and $\hat{d}|\tilde{\alpha}_{2}\rangle=0$, respectively. Applying
$\hat{d}$ to the above resolution of $|\tilde{\alpha}\rangle$ yields
$|\alpha\rangle=\sum_{\tilde{\alpha}_{1}}C_{\tilde{\alpha}_{1}}\hat{d}|\tilde{\alpha}_{1}\rangle$.
By the supersymmetry of $\hat{H}$, any $\hat{d}|\tilde{\alpha}_{1}\rangle$
is an eigenstates of $\hat{H}^{(k)}$ with the same eigenvalue as
$|\tilde{\alpha}_{1}\rangle$. Furthermore, in the above resolution
of $|\alpha\rangle$ only eigenstates with the same eigenvalue as
$|\alpha\rangle$ has itself can appear. This is because each eigenstate
of a complete basis is linearly independent of the other eigenstates
and in particular cannot be given as a linear combination of other
eigenstates with different eigenvalues. Thus, in the most general
situation, when there are no additional degeneracies, one eigenstate
$|\tilde{\alpha}'\rangle$ of $\hat{H}^{(k-1)}$ must exist such that
$|\alpha\rangle=\hat{d}|\tilde{\alpha}'\rangle$ and $|\tilde{\alpha}\rangle=|\tilde{\alpha}'\rangle+|\tilde{\alpha}''\rangle$,
with $\hat{d}|\tilde{\alpha}''\rangle=0$.}}\textcolor{black}{{} Again, we have a boson-fermion pair of eigenstates
$|\alpha\rangle=\hat{d}|\tilde{\alpha}'\rangle\in\Omega^{k}$ and
$|\tilde{\alpha}'\rangle\in\Omega^{k-1}$ of the same eigenvalue.}
\end{enumerate}
In this manner, all eigenstates with non-zero eigenvalues are non-$\hat{d}$-symmetric
boson-fermion pairs of states. Thus we just came to an important conclusion
-- all supersymmetric singlet states have vanishing eigenvalues. 

This implies that any growing mode of a KD cannot be a supersymmetric
state of the corresponding SEO. Operational KDs must exhibit a broken
supersymmetry.

\subsection{Non-magnetic modes}

Let us now investigate whether the dynamo phenomenon must be intrinsic
to magnetic fields, or whether other quantities described by 0- or
3-forms could be amplified exponentially by the Lie action of a stationary
flow field. 

\textcolor{black}{Any physically meaningful stochastic model must
have the steady-state (zero-eigenvalue) total probability distribution,}\footnote{\textcolor{black}{In the theory of deterministic dynamics, the counterpart
of this state is known as invariant measure and the Krylov\textendash Bogolyubov
theorem states that it always exists.}}\textcolor{black}{{} $\psi_{\mathrm{TE}}=P_{\mathrm{TE}}d^{3}x\in\Omega^{3}\:$
with $\int_{x}\psi_{\mathrm{TE}}=1.$ This state can be recognized
as the state of thermodynamic equilibrium (TE). This eigenstate is
supersymmetric because $\hat{d}\psi_{\mathrm{TE}}=0$, which is true
for all 3-forms, and it is not $\hat{d}$-exact because otherwise
its integral over $X$ would vanish. This state is always the ``ground
state'' for $\Omega^{3}$. In other words, among all the eigenstates
in $\Omega^{3}$, the state of thermodynamic equilibrium has the least
real part of its eigenvalue and this is zero. This can be seen from
the following qualitative, yet robust argument. }

\textcolor{black}{All non-$\hat{d}$-symmetric eigenstates $\alpha$
from $\Omega^{3}$ are $\hat{d}$-exact, i.e., of the form $|\alpha\rangle=\hat{d}|\alpha'\rangle$.}\footnote{\textcolor{black}{All non-$\hat{d}$-symmetric pairs of states are
of the form $|\alpha'\rangle$ and $|\alpha\rangle=\hat{d}|\alpha'\rangle\ne0$.
Since all 3-forms are $\hat{d}$-closed, the non-$\hat{d}$-symmetric
3-forms can only be of the type $|\alpha\rangle=\hat{d}|\alpha'\rangle$.}}\textcolor{black}{. The integral of the kets of these states is zero,
$\int_{X}\alpha=\int_{X}\hat{d}\alpha'=0$, which means that their
wavefunctions must be negative at least somewhere on $X$. }

\textcolor{black}{Imagine now that the ``ground state'' on $\Omega^{3}$
is a non-$\hat{d}$-symmetric eigenstate with a real negative eigenvalue.
An arbitrary total probability distribution will have non-zero contribution
from such an eigenstate. Its temporal evolution according to Eq.~\eqref{eq:LinearEvolution}
will lead to a state dominated by this eigenstate after sufficiently
long time of evolution as this state is an exponentially growing mode.
Consequently, the total probability distribution will become negative
somewhere on the phase space, since this state must be negative in
some regions of the phase space as discussed in the previous paragraph.}

\textcolor{black}{A negative probability distribution is of course
}illogical\textcolor{black}{{} and therefore in contradiction to the
above assumption that a non-$\hat{d}$-symmetric state with negative
eigenvalue exists in $\Omega^{3}$. Using similar reasoning one can
also rule out the possibility of a pair of Ruelle-Pollicott resonances
with a negative real part of their eigenvalues being the ground states
on $\Omega^{3}$. Thus, one arrives at the conclusion that the supersymmetric
state of thermodynamic equilibrium is always the ``ground state''
of $\Omega^{3}$. It can be said that the conventional Focker-Planck
operator $\hat{H}^{(3)}$ never breaks supersymmetry spontaneously,
i.e., its ground state is always supersymmetric.}

This means that density modes do not grow infinitely under a stationary
flow, but saturate in this supersymmetric state.\footnote{The cosmologically educated reader might wonder whether the growing
modes of cosmic structure formation are not a counter example. They
are not, as these appear in a linearized description of the transport
of cosmic matter. Once all matter is swept into cosmic structures
(in a non-expanding Universe, to make the mathematical analogy fit)
it will stay there. These structures cannot grow further once their
supply regions are empty. Merging of structures would require a temporal
change of the velocity field, which is outside the class of systems
investigated in this work.}

\textcolor{black}{It can also be shown that $\hat{H}^{(0)}$ is related
by a similarity transformation to $\hat{H}_{\mathrm{T}}^{(3)}$, where
$\hat{H}_{\mathrm{T}}$ is the SEO of the time-reversed SDE, i.e.,
the SDE with the opposite flow vector field and the above argumentation
also applies to $\hat{H}_{\mathrm{T}}^{(3)}$. Operators related by
a similarity transformation are isospectral so that $\hat{H}^{(0)}$
must also never break the supersymmetry on its own, just as $\hat{H}^{(3)}$.}\footnote{\textcolor{black}{The supersymmetric ``ground state'' of $\hat{H}^{(0)}$
is a constant function on $X$. }}\textcolor{black}{{} In conclusion, in three dimensions, only non-supersymmetric
1-forms and 2-forms can spontaneously break the overall supersymmetry
of the SEO. For this reason, a dynamo mechanism is only known for
the vector potential of the magnetic fields, but not for the scalar
potential of the electric field. }

The above discussion leads to the conclusion that there are only three
possible forms of spectra of SEO (see Fig.~\ref{fig:The-three-possible}),
with the 1- and 2-forms being the eigenstates of the KD operator.
The first type in Fig.~1a corresponds to the unbroken topological
supersymmetry because the ground states of the model are supersymmetric.\footnote{By coincidence, non-supersymmetric eigenstates with zero eigenvalues
can exist. Such degeneracy can be removed by a deformation of the
model.} The other two types of spectra correspond to the spontaneously broken
topological supersymmetry because the ground states have non-zero
eigenvalues and therefore are non-supersymmetric. These correspond
to growing KD modes, either just exponentially growing, or with a
superimposed oscillation.

\subsection{Stochastic chaos}

It is known that non-diffusive KDs require some non-integrability
of the underlying flow, which is a signature of (deterministic) chaos
\citep{ChaosDynamo}.The centerpiece of the theory of deterministic
chaos is the butterfly effect \textcolor{black}{(BE),} which is a
high sensitivity of the subsequent evolution of a system to perturbations
and/or variations in initial conditions. The BE is often regarded
as the defining property of chaos \citep{Motter_Chaos,Ruelle_Chaos}.
The supersymmetry breaking picture of stochastic chaos provides a
theoretical explanation of the BE via the Goldstone theorem \citep{Mine_2}.

In order to identify spontaneous topological supersymmetry breaking
with the emergence of deterministic chaos we investigate the number
of periodic trajectories of a dynamical systems. It is well-known
that in many deterministic chaotic systems, the number of periodic
trajectories grows exponentially as a function of their periods in
the long-time limit.\footnote{The exponential rate of this growth is related to various versions
of entropy (such as topological entropy, see, e.g., Ref.\citep{BookEntropy}
and refs therein) introduced in dynamical sytsem theory.} This exponential growth comes from the infinite number of unstable
periodic orbits with unlimited periods that constitute strange or
chaotic attractors \citep{Gilmore}. In Ref.~\citep{Mine_2}, it
was shown that for some classes of models the stochastically averaged
number of periodic solutions must be represented by the dynamical
partition function,

\begin{eqnarray*}
Z_{t}|_{t\to\infty}=\mathrm{Tr}\,e^{-t\hat{H}}|_{t\to\infty}\approx
\end{eqnarray*}

\begin{eqnarray*}
\approx\sum_{g}e^{-t\mathcal{E}_{g}}=\begin{cases}
\begin{array}{cc}
\mathrm{const} & \mbox{Fig. 1a,}\\
2\,e^{t\,|\mathcal{E}_{g}|} & \mbox{Fig. 1b,}\\
4\,\cos(t\,\mathrm{Im}\mathcal{E}_{g})\,e^{t\,|\mathrm{Re}\mathcal{E}_{g}|} & \mbox{Fig. 1c,}
\end{array}\end{cases}.
\end{eqnarray*}
where the label \textcolor{blue}{$g$ }runs over the ground states.
This equation shows that for spectra as displayed in Fig.~1b, the
stochastically averaged number of periodic solutions must grow exponentially.
The broken topological supersymmetry therefore seems to be a prerequisite
for deterministic chaos in deterministic as well as in stochastic
systems. Therefore, this spectral classification of dynamical chaos
seems to be the natural generalization of the concept of deterministic
chaos to the stochastic regime.

In the case of a spectrum as displayed in Fig.~1c, the dynamical
partition function can become negative. This implies that in this
case the dynamical partition function cannot represent the stochastically
averaged number of periodic solutions. Furthermore, there are theorems
in the dynamical system theory stating that for a certain class of
(expanding) models that mimic chaotic behavior, the eigenvalue of
the ground state must be real.\footnote{In terminology of Ref.\citep{Rue02}, the finite-time SEO is the generalized
transfer operator (GTO), and it is the spectrum of the GTO which is
addressed there.} In the light of these theorems, the spectrum in Fig.~1c looks suspicious. 

This raises the question whether such STS spectra are realizable.\textcolor{black}{{}
Here, the KD-STS correspondence proves useful for the STS. As flow
configurations which exhibit growing oscillating KD modes are known
}\citep{1987AN....308...89B,1988JFM...197...39R,1220.78120}\textcolor{black}{{}
the existence of spectra like shown in Fig.}~\textcolor{black}{1c
are realizable. }

\subsection{The limit of the KD-STS correspondence}

The reported KD-STS correspondence is in a sense accidental. It is
possible due to the linearity of the induction equation and its specific
form, i.e., the Lie derivative along the velocity field representing
the evolution of magnetic field lines frozen into the media. In a
more general situation as given by the nonlinear dynamo, this picture
breaks down. On the other hand, the STS is applicable to all stochastic
and deterministic differential equations and the dynamical equations
of the non-linear dynamo is one of them. If the STS theory of the
nonlinear dynamo would be constructed, it would be a full-fledged
``field theory'' with an infinite-dimensional phase space of magnetic
field configurations as compared to the finite-dimensional phase space
of test particle trajectories considered here for the KD. Instead
of Eq.~\eqref{eq:CorrespondingSDEREd}, the underlying equation of
motion will be the induction equation itself. The magnetic field (and
perhaps other functions/fields over the real space) would no longer
be the ``wavefunction'', as in case of the KD, but rather the coordinates
of the model, whereas the wavefunctions would depend on them in a
functional manner.

\section{Conclusion\label{sec:Conclusion} }

We established a correspondence between the theory of KD and the recently
found approximation-free STS. We showed that the KD equation is essentially
the stochastic evolution operator of a related SDE and thus it possesses
topological supersymmetry. This connection allowed us to identify
the KD effect as the supersymmetry breaking phenomenon in the corresponding
SDE. We further argued that the SDEs related to KDs with growing magnetic
modes are chaotic in a generalized stochastic sense. We showed that
the flow pattern of stationary KDs can only amplify quantities associated
with vector potentials (like magnetic fields) but not those associated
with scalar potentials or which are densities. 

Furthermore, the existence of the growing modes of the magnetic field
in the KDs can be viewed as a proof that the supersymmetry breaking
spectra of the stochastic evolution operator with both real and complex
conjugate eigenvalues of the ground states are realizable. This finding
is valuable for the STS. We believe that further work may reveal other
important insights on both sides of the KD-STS correspondence established
in this paper.

\section{Acknowledgements}

This work was supported partly by the Excellence Cluster Universe.
I.V.O. would like to thank the Max Planck Institute for Astrophysics
for the hospitality during his visit in the summer 2015. T.A.E. thanks
the International Space Science Institute (ISSI) in Bern for its hospitality
and the ISSI International Team 323 for a stimulating atmosphere.
We acknowledge valuable comments on the manuscript by David Butler,
Sebastian Dorn, Maksim Greiner, Reimar Leike, Daniel Pumpe and Theo
Steininger.

\bibliographystyle{apsrev4-1}
\bibliography{tft,dynamo}

\begin{thebibliography}{48}%
\makeatletter
\providecommand \@ifxundefined [1]{%
 \@ifx{#1\undefined}
}%
\providecommand \@ifnum [1]{%
 \ifnum #1\expandafter \@firstoftwo
 \else \expandafter \@secondoftwo
 \fi
}%
\providecommand \@ifx [1]{%
 \ifx #1\expandafter \@firstoftwo
 \else \expandafter \@secondoftwo
 \fi
}%
\providecommand \natexlab [1]{#1}%
\providecommand \enquote  [1]{``#1''}%
\providecommand \bibnamefont  [1]{#1}%
\providecommand \bibfnamefont [1]{#1}%
\providecommand \citenamefont [1]{#1}%
\providecommand \href@noop [0]{\@secondoftwo}%
\providecommand \href [0]{\begingroup \@sanitize@url \@href}%
\providecommand \@href[1]{\@@startlink{#1}\@@href}%
\providecommand \@@href[1]{\endgroup#1\@@endlink}%
\providecommand \@sanitize@url [0]{\catcode `\\12\catcode `\$12\catcode
  `\&12\catcode `\#12\catcode `\^12\catcode `\_12\catcode `\%12\relax}%
\providecommand \@@startlink[1]{}%
\providecommand \@@endlink[0]{}%
\providecommand \url  [0]{\begingroup\@sanitize@url \@url }%
\providecommand \@url [1]{\endgroup\@href {#1}{\urlprefix }}%
\providecommand \urlprefix  [0]{URL }%
\providecommand \Eprint [0]{\href }%
\providecommand \doibase [0]{http://dx.doi.org/}%
\providecommand \selectlanguage [0]{\@gobble}%
\providecommand \bibinfo  [0]{\@secondoftwo}%
\providecommand \bibfield  [0]{\@secondoftwo}%
\providecommand \translation [1]{[#1]}%
\providecommand \BibitemOpen [0]{}%
\providecommand \bibitemStop [0]{}%
\providecommand \bibitemNoStop [0]{.\EOS\space}%
\providecommand \EOS [0]{\spacefactor3000\relax}%
\providecommand \BibitemShut  [1]{\csname bibitem#1\endcsname}%
\let\auto@bib@innerbib\@empty
\bibitem [{\citenamefont {{Bullard}}\ and\ \citenamefont
  {{Gellman}}(1954)}]{1954RSPTA.247..213B}%
  \BibitemOpen
  \bibfield  {author} {\bibinfo {author} {\bibfnamefont {E.}~\bibnamefont
  {{Bullard}}}\ and\ \bibinfo {author} {\bibfnamefont {H.}~\bibnamefont
  {{Gellman}}},\ }\href {\doibase 10.1098/rsta.1954.0018} {\bibfield  {journal}
  {\bibinfo  {journal} {Royal Society of London Philosophical Transactions
  Series A}\ }\textbf {\bibinfo {volume} {247}},\ \bibinfo {pages} {213}
  (\bibinfo {year} {1954})}\BibitemShut {NoStop}%
\bibitem [{\citenamefont {{Parker}}(1955)}]{1955ApJ...122..293P}%
  \BibitemOpen
  \bibfield  {author} {\bibinfo {author} {\bibfnamefont {E.~N.}\ \bibnamefont
  {{Parker}}},\ }\href {\doibase 10.1086/146087} {\bibfield  {journal}
  {\bibinfo  {journal} {\apj}\ }\textbf {\bibinfo {volume} {122}},\ \bibinfo
  {pages} {293} (\bibinfo {year} {1955})}\BibitemShut {NoStop}%
\bibitem [{\citenamefont {{Braginskiy}}(1964)}]{1964Ge&Ae...4..572B}%
  \BibitemOpen
  \bibfield  {author} {\bibinfo {author} {\bibfnamefont {S.~I.}\ \bibnamefont
  {{Braginskiy}}},\ }\href@noop {} {\bibfield  {journal} {\bibinfo  {journal}
  {Geomagnetism and Aeronomy}\ }\textbf {\bibinfo {volume} {4}},\ \bibinfo
  {pages} {572} (\bibinfo {year} {1964})}\BibitemShut {NoStop}%
\bibitem [{\citenamefont {{Parker}}(1970)}]{1970ApJ...162..665P}%
  \BibitemOpen
  \bibfield  {author} {\bibinfo {author} {\bibfnamefont {E.~N.}\ \bibnamefont
  {{Parker}}},\ }\href {\doibase 10.1086/150697} {\bibfield  {journal}
  {\bibinfo  {journal} {\apj}\ }\textbf {\bibinfo {volume} {162}},\ \bibinfo
  {pages} {665} (\bibinfo {year} {1970})}\BibitemShut {NoStop}%
\bibitem [{\citenamefont {{Krause}}\ and\ \citenamefont
  {{Raedler}}(1980)}]{1980opp..bookR....K}%
  \BibitemOpen
  \bibfield  {author} {\bibinfo {author} {\bibfnamefont {F.}~\bibnamefont
  {{Krause}}}\ and\ \bibinfo {author} {\bibfnamefont {K.-H.}\ \bibnamefont
  {{Raedler}}},\ }\href@noop {} {\emph {\bibinfo {title} {Organic Photonics and
  Photovoltaics}}}\ (\bibinfo {year} {1980})\BibitemShut {NoStop}%
\bibitem [{\citenamefont {Rüdiger}(2004)}]{Book_on_Magnetic_Dynamo}%
  \BibitemOpen
  \bibfield  {author} {\bibinfo {author} {\bibfnamefont {R.}~\bibnamefont
  {Rüdiger}, \bibfnamefont {G.~Hollerbach}},\ }\href@noop {} {\emph {\bibinfo
  {title} {The Magnetic Universe: Geophysical and Astrophysical Dynamo
  Theory}}}\ (\bibinfo  {publisher} {Wiley-VCH Verlag GmbH \& Co, KGaA},\
  \bibinfo {year} {2004})\BibitemShut {NoStop}%
\bibitem [{\citenamefont {{Parker}}(1992)}]{1992ApJ...401..137P}%
  \BibitemOpen
  \bibfield  {author} {\bibinfo {author} {\bibfnamefont {E.~N.}\ \bibnamefont
  {{Parker}}},\ }\href {\doibase 10.1086/172046} {\bibfield  {journal}
  {\bibinfo  {journal} {\apj}\ }\textbf {\bibinfo {volume} {401}},\ \bibinfo
  {pages} {137} (\bibinfo {year} {1992})}\BibitemShut {NoStop}%
\bibitem [{\citenamefont {{Beck}}\ \emph {et~al.}(1994)\citenamefont {{Beck}},
  \citenamefont {{Poezd}}, \citenamefont {{Shukurov}},\ and\ \citenamefont
  {{Sokoloff}}}]{1994A&A...289...94B}%
  \BibitemOpen
  \bibfield  {author} {\bibinfo {author} {\bibfnamefont {R.}~\bibnamefont
  {{Beck}}}, \bibinfo {author} {\bibfnamefont {A.~D.}\ \bibnamefont {{Poezd}}},
  \bibinfo {author} {\bibfnamefont {A.}~\bibnamefont {{Shukurov}}}, \ and\
  \bibinfo {author} {\bibfnamefont {D.~D.}\ \bibnamefont {{Sokoloff}}},\
  }\href@noop {} {\bibfield  {journal} {\bibinfo  {journal} {\aap}\ }\textbf
  {\bibinfo {volume} {289}},\ \bibinfo {pages} {94} (\bibinfo {year}
  {1994})}\BibitemShut {NoStop}%
\bibitem [{\citenamefont {{Subramanian}}(1998)}]{1998MNRAS.294..718S}%
  \BibitemOpen
  \bibfield  {author} {\bibinfo {author} {\bibfnamefont {K.}~\bibnamefont
  {{Subramanian}}},\ }\href {\doibase 10.1046/j.1365-8711.1998.01284.x}
  {\bibfield  {journal} {\bibinfo  {journal} {\mnras}\ }\textbf {\bibinfo
  {volume} {294}},\ \bibinfo {pages} {718} (\bibinfo {year} {1998})},\ \Eprint
  {http://arxiv.org/abs/astro-ph/9707280} {astro-ph/9707280} \BibitemShut
  {NoStop}%
\bibitem [{\citenamefont {{Beck}}(2007)}]{2007A&A...470..539B}%
  \BibitemOpen
  \bibfield  {author} {\bibinfo {author} {\bibfnamefont {R.}~\bibnamefont
  {{Beck}}},\ }\href {\doibase 10.1051/0004-6361:20066988} {\bibfield
  {journal} {\bibinfo  {journal} {\aap}\ }\textbf {\bibinfo {volume} {470}},\
  \bibinfo {pages} {539} (\bibinfo {year} {2007})},\ \Eprint
  {http://arxiv.org/abs/0705.4163} {arXiv:0705.4163} \BibitemShut {NoStop}%
\bibitem [{\citenamefont {{Ruzmaikin}}\ \emph {et~al.}(1989)\citenamefont
  {{Ruzmaikin}}, \citenamefont {{Sokolov}},\ and\ \citenamefont
  {{Shukurov}}}]{1989MNRAS.241....1R}%
  \BibitemOpen
  \bibfield  {author} {\bibinfo {author} {\bibfnamefont {A.}~\bibnamefont
  {{Ruzmaikin}}}, \bibinfo {author} {\bibfnamefont {D.}~\bibnamefont
  {{Sokolov}}}, \ and\ \bibinfo {author} {\bibfnamefont {A.}~\bibnamefont
  {{Shukurov}}},\ }\href@noop {} {\bibfield  {journal} {\bibinfo  {journal}
  {\mnras}\ }\textbf {\bibinfo {volume} {241}},\ \bibinfo {pages} {1} (\bibinfo
  {year} {1989})}\BibitemShut {NoStop}%
\bibitem [{\citenamefont {{Dolag}}\ \emph {et~al.}(2002)\citenamefont
  {{Dolag}}, \citenamefont {{Bartelmann}},\ and\ \citenamefont
  {{Lesch}}}]{2002A&A...387..383D}%
  \BibitemOpen
  \bibfield  {author} {\bibinfo {author} {\bibfnamefont {K.}~\bibnamefont
  {{Dolag}}}, \bibinfo {author} {\bibfnamefont {M.}~\bibnamefont
  {{Bartelmann}}}, \ and\ \bibinfo {author} {\bibfnamefont {H.}~\bibnamefont
  {{Lesch}}},\ }\href {\doibase 10.1051/0004-6361:20020241} {\bibfield
  {journal} {\bibinfo  {journal} {\aap}\ }\textbf {\bibinfo {volume} {387}},\
  \bibinfo {pages} {383} (\bibinfo {year} {2002})}\BibitemShut {NoStop}%
\bibitem [{\citenamefont {{En{\ss}lin}}\ and\ \citenamefont
  {{Vogt}}(2006)}]{2006A&A...453..447E}%
  \BibitemOpen
  \bibfield  {author} {\bibinfo {author} {\bibfnamefont {T.~A.}\ \bibnamefont
  {{En{\ss}lin}}}\ and\ \bibinfo {author} {\bibfnamefont {C.}~\bibnamefont
  {{Vogt}}},\ }\href {\doibase 10.1051/0004-6361:20053518} {\bibfield
  {journal} {\bibinfo  {journal} {\aap}\ }\textbf {\bibinfo {volume} {453}},\
  \bibinfo {pages} {447} (\bibinfo {year} {2006})},\ \Eprint
  {http://arxiv.org/abs/astro-ph/0505517} {astro-ph/0505517} \BibitemShut
  {NoStop}%
\bibitem [{\citenamefont {{Vazza}}\ \emph {et~al.}(2014)\citenamefont
  {{Vazza}}, \citenamefont {{Br{\"u}ggen}}, \citenamefont {{Gheller}},\ and\
  \citenamefont {{Wang}}}]{2014MNRAS.445.3706V}%
  \BibitemOpen
  \bibfield  {author} {\bibinfo {author} {\bibfnamefont {F.}~\bibnamefont
  {{Vazza}}}, \bibinfo {author} {\bibfnamefont {M.}~\bibnamefont
  {{Br{\"u}ggen}}}, \bibinfo {author} {\bibfnamefont {C.}~\bibnamefont
  {{Gheller}}}, \ and\ \bibinfo {author} {\bibfnamefont {P.}~\bibnamefont
  {{Wang}}},\ }\href {\doibase 10.1093/mnras/stu1896} {\bibfield  {journal}
  {\bibinfo  {journal} {\mnras}\ }\textbf {\bibinfo {volume} {445}},\ \bibinfo
  {pages} {3706} (\bibinfo {year} {2014})},\ \Eprint
  {http://arxiv.org/abs/1409.2640} {arXiv:1409.2640} \BibitemShut {NoStop}%
\bibitem [{\citenamefont {{Parker}}(1975)}]{1975ApJ...198..205P}%
  \BibitemOpen
  \bibfield  {author} {\bibinfo {author} {\bibfnamefont {E.~N.}\ \bibnamefont
  {{Parker}}},\ }\href {\doibase 10.1086/153593} {\bibfield  {journal}
  {\bibinfo  {journal} {\apj}\ }\textbf {\bibinfo {volume} {198}},\ \bibinfo
  {pages} {205} (\bibinfo {year} {1975})}\BibitemShut {NoStop}%
\bibitem [{\citenamefont {{Glatzmaier}}(1984)}]{1984JCoPh..55..461G}%
  \BibitemOpen
  \bibfield  {author} {\bibinfo {author} {\bibfnamefont {G.~A.}\ \bibnamefont
  {{Glatzmaier}}},\ }\href {\doibase 10.1016/0021-9991(84)90033-0} {\bibfield
  {journal} {\bibinfo  {journal} {Journal of Computational Physics}\ }\textbf
  {\bibinfo {volume} {55}},\ \bibinfo {pages} {461} (\bibinfo {year}
  {1984})}\BibitemShut {NoStop}%
\bibitem [{\citenamefont {{Raedler}}(1986)}]{1986AN....307...89R}%
  \BibitemOpen
  \bibfield  {author} {\bibinfo {author} {\bibfnamefont {K.-H.}\ \bibnamefont
  {{Raedler}}},\ }\href {\doibase 10.1002/asna.2113070205} {\bibfield
  {journal} {\bibinfo  {journal} {Astronomische Nachrichten}\ }\textbf
  {\bibinfo {volume} {307}},\ \bibinfo {pages} {89} (\bibinfo {year}
  {1986})}\BibitemShut {NoStop}%
\bibitem [{\citenamefont {{Charbonneau}}\ and\ \citenamefont
  {{MacGregor}}(2001)}]{2001ApJ...559.1094C}%
  \BibitemOpen
  \bibfield  {author} {\bibinfo {author} {\bibfnamefont {P.}~\bibnamefont
  {{Charbonneau}}}\ and\ \bibinfo {author} {\bibfnamefont {K.~B.}\ \bibnamefont
  {{MacGregor}}},\ }\href {\doibase 10.1086/322417} {\bibfield  {journal}
  {\bibinfo  {journal} {\apj}\ }\textbf {\bibinfo {volume} {559}},\ \bibinfo
  {pages} {1094} (\bibinfo {year} {2001})}\BibitemShut {NoStop}%
\bibitem [{\citenamefont {{Spruit}}(2002)}]{2002A&A...381..923S}%
  \BibitemOpen
  \bibfield  {author} {\bibinfo {author} {\bibfnamefont {H.~C.}\ \bibnamefont
  {{Spruit}}},\ }\href {\doibase 10.1051/0004-6361:20011465} {\bibfield
  {journal} {\bibinfo  {journal} {\aap}\ }\textbf {\bibinfo {volume} {381}},\
  \bibinfo {pages} {923} (\bibinfo {year} {2002})},\ \Eprint
  {http://arxiv.org/abs/astro-ph/0108207} {astro-ph/0108207} \BibitemShut
  {NoStop}%
\bibitem [{\citenamefont {{Braithwaite}}(2006)}]{2006A&A...449..451B}%
  \BibitemOpen
  \bibfield  {author} {\bibinfo {author} {\bibfnamefont {J.}~\bibnamefont
  {{Braithwaite}}},\ }\href {\doibase 10.1051/0004-6361:20054241} {\bibfield
  {journal} {\bibinfo  {journal} {\aap}\ }\textbf {\bibinfo {volume} {449}},\
  \bibinfo {pages} {451} (\bibinfo {year} {2006})},\ \Eprint
  {http://arxiv.org/abs/astro-ph/0509693} {astro-ph/0509693} \BibitemShut
  {NoStop}%
\bibitem [{\citenamefont {{Browning}}(2008)}]{2008ApJ...676.1262B}%
  \BibitemOpen
  \bibfield  {author} {\bibinfo {author} {\bibfnamefont {M.~K.}\ \bibnamefont
  {{Browning}}},\ }\href {\doibase 10.1086/527432} {\bibfield  {journal}
  {\bibinfo  {journal} {\apj}\ }\textbf {\bibinfo {volume} {676}},\ \bibinfo
  {pages} {1262} (\bibinfo {year} {2008})},\ \Eprint
  {http://arxiv.org/abs/0712.1603} {arXiv:0712.1603} \BibitemShut {NoStop}%
\bibitem [{\citenamefont {{Taylor}}(1963)}]{1963RSPSA.274..274T}%
  \BibitemOpen
  \bibfield  {author} {\bibinfo {author} {\bibfnamefont {J.~B.}\ \bibnamefont
  {{Taylor}}},\ }\href {\doibase 10.1098/rspa.1963.0130} {\bibfield  {journal}
  {\bibinfo  {journal} {Royal Society of London Proceedings Series A}\ }\textbf
  {\bibinfo {volume} {274}},\ \bibinfo {pages} {274} (\bibinfo {year}
  {1963})}\BibitemShut {NoStop}%
\bibitem [{\citenamefont {{Kuang}}\ and\ \citenamefont
  {{Bloxham}}(1997)}]{1997Natur.389..371K}%
  \BibitemOpen
  \bibfield  {author} {\bibinfo {author} {\bibfnamefont {W.}~\bibnamefont
  {{Kuang}}}\ and\ \bibinfo {author} {\bibfnamefont {J.}~\bibnamefont
  {{Bloxham}}},\ }\href {\doibase 10.1038/38712} {\bibfield  {journal}
  {\bibinfo  {journal} {\nat}\ }\textbf {\bibinfo {volume} {389}},\ \bibinfo
  {pages} {371} (\bibinfo {year} {1997})}\BibitemShut {NoStop}%
\bibitem [{\citenamefont {{Lerche}}(1971)}]{1971ApJ...166..627L}%
  \BibitemOpen
  \bibfield  {author} {\bibinfo {author} {\bibfnamefont {I.}~\bibnamefont
  {{Lerche}}},\ }\href {\doibase 10.1086/150988} {\bibfield  {journal}
  {\bibinfo  {journal} {\apj}\ }\textbf {\bibinfo {volume} {166}},\ \bibinfo
  {pages} {627} (\bibinfo {year} {1971})}\BibitemShut {NoStop}%
\bibitem [{\citenamefont {{Roberts}}(1972)}]{1972RSPTA.272..663R}%
  \BibitemOpen
  \bibfield  {author} {\bibinfo {author} {\bibfnamefont {P.~H.}\ \bibnamefont
  {{Roberts}}},\ }\href {\doibase 10.1098/rsta.1972.0074} {\bibfield  {journal}
  {\bibinfo  {journal} {Royal Society of London Philosophical Transactions
  Series A}\ }\textbf {\bibinfo {volume} {272}},\ \bibinfo {pages} {663}
  (\bibinfo {year} {1972})}\BibitemShut {NoStop}%
\bibitem [{\citenamefont {Li}\ \emph {et~al.}(2010{\natexlab{a}})\citenamefont
  {Li}, \citenamefont {Livermore},\ and\ \citenamefont
  {Jackson}}]{Dynamo_Complex}%
  \BibitemOpen
  \bibfield  {author} {\bibinfo {author} {\bibfnamefont {K.}~\bibnamefont
  {Li}}, \bibinfo {author} {\bibfnamefont {P.~W.}\ \bibnamefont {Livermore}}, \
  and\ \bibinfo {author} {\bibfnamefont {A.}~\bibnamefont {Jackson}},\
  }\href@noop {} {\bibfield  {journal} {\bibinfo  {journal} {Journal of
  Computational Physics}\ }\textbf {\bibinfo {volume} {229}},\ \bibinfo {pages}
  {8666} (\bibinfo {year} {2010}{\natexlab{a}})}\BibitemShut {NoStop}%
\bibitem [{\citenamefont {Motter}(2014)}]{Motter_Chaos}%
  \BibitemOpen
  \bibfield  {author} {\bibinfo {author} {\bibfnamefont {D.~K.}\ \bibnamefont
  {Motter}, \bibfnamefont {Adilson E.~Campbell}},\ }\href@noop {} {\bibfield
  {journal} {\bibinfo  {journal} {Physics Today}\ }\textbf {\bibinfo {volume}
  {66}},\ \bibinfo {pages} {27} (\bibinfo {year} {2014})}\BibitemShut {NoStop}%
\bibitem [{\citenamefont {Ott}(2007)}]{ChaosDynamo}%
  \BibitemOpen
  \bibfield  {author} {\bibinfo {author} {\bibfnamefont {E.}~\bibnamefont
  {Ott}},\ }\href@noop {} {\bibfield  {journal} {\bibinfo  {journal} {Chaos,
  Kinetics and Nonlinear Dynamics in Fluids and Plasmas}\ }\textbf {\bibinfo
  {volume} {511}},\ \bibinfo {pages} {263} (\bibinfo {year}
  {2007})}\BibitemShut {NoStop}%
\bibitem [{\citenamefont {Ovchinnikov}(2013)}]{Mine_1}%
  \BibitemOpen
  \bibfield  {author} {\bibinfo {author} {\bibfnamefont {I.~V.}\ \bibnamefont
  {Ovchinnikov}},\ }\href@noop {} {\bibfield  {journal} {\bibinfo  {journal}
  {Chaos: An Interdisciplinary Journal of Nonlinear Science}\ }\textbf
  {\bibinfo {volume} {23}},\ \bibinfo {pages} {013108} (\bibinfo {year}
  {2013})}\BibitemShut {NoStop}%
\bibitem [{\citenamefont {Ovchinnikov}(2014)}]{Mine_2}%
  \BibitemOpen
  \bibfield  {author} {\bibinfo {author} {\bibfnamefont {I.~V.}\ \bibnamefont
  {Ovchinnikov}},\ }\href@noop {} {\bibfield  {journal} {\bibinfo  {journal}
  {arXiv:1308.4222}\ } (\bibinfo {year} {2014})}\BibitemShut {NoStop}%
\bibitem [{\citenamefont {{Baryshnikova}}\ and\ \citenamefont
  {{Shukurov}}(1987)}]{1987AN....308...89B}%
  \BibitemOpen
  \bibfield  {author} {\bibinfo {author} {\bibfnamefont {I.}~\bibnamefont
  {{Baryshnikova}}}\ and\ \bibinfo {author} {\bibfnamefont {A.}~\bibnamefont
  {{Shukurov}}},\ }\href {\doibase 10.1002/asna.2113080202} {\bibfield
  {journal} {\bibinfo  {journal} {Astronomische Nachrichten}\ }\textbf
  {\bibinfo {volume} {308}},\ \bibinfo {pages} {89} (\bibinfo {year}
  {1987})}\BibitemShut {NoStop}%
\bibitem [{\citenamefont {{Ruzmaikin}}\ \emph {et~al.}(1988)\citenamefont
  {{Ruzmaikin}}, \citenamefont {{Sokolov}},\ and\ \citenamefont
  {{Shukurov}}}]{1988JFM...197...39R}%
  \BibitemOpen
  \bibfield  {author} {\bibinfo {author} {\bibfnamefont {A.}~\bibnamefont
  {{Ruzmaikin}}}, \bibinfo {author} {\bibfnamefont {D.}~\bibnamefont
  {{Sokolov}}}, \ and\ \bibinfo {author} {\bibfnamefont {A.}~\bibnamefont
  {{Shukurov}}},\ }\href {\doibase 10.1017/S0022112088003167} {\bibfield
  {journal} {\bibinfo  {journal} {Journal of Fluid Mechanics}\ }\textbf
  {\bibinfo {volume} {197}},\ \bibinfo {pages} {39} (\bibinfo {year}
  {1988})}\BibitemShut {NoStop}%
\bibitem [{\citenamefont {Li}\ \emph {et~al.}(2010{\natexlab{b}})\citenamefont
  {Li}, \citenamefont {Livermore},\ and\ \citenamefont {Jackson}}]{1220.78120}%
  \BibitemOpen
  \bibfield  {author} {\bibinfo {author} {\bibfnamefont {K.}~\bibnamefont
  {Li}}, \bibinfo {author} {\bibfnamefont {P.~W.}\ \bibnamefont {Livermore}}, \
  and\ \bibinfo {author} {\bibfnamefont {A.}~\bibnamefont {Jackson}},\ }\href
  {\doibase 10.1016/j.jcp.2010.07.039} {\bibfield  {journal} {\bibinfo
  {journal} {J. Comput. Phys.}\ }\textbf {\bibinfo {volume} {229}},\ \bibinfo
  {pages} {8666} (\bibinfo {year} {2010}{\natexlab{b}})}\BibitemShut {NoStop}%
\bibitem [{\citenamefont {Dutta}\ and\ \citenamefont
  {Horn}(1981)}]{RevModPhys.53.497}%
  \BibitemOpen
  \bibfield  {author} {\bibinfo {author} {\bibfnamefont {P.}~\bibnamefont
  {Dutta}}\ and\ \bibinfo {author} {\bibfnamefont {P.~M.}\ \bibnamefont
  {Horn}},\ }\href {\doibase 10.1103/RevModPhys.53.497} {\bibfield  {journal}
  {\bibinfo  {journal} {Rev. Mod. Phys.}\ }\textbf {\bibinfo {volume} {53}},\
  \bibinfo {pages} {497} (\bibinfo {year} {1981})}\BibitemShut {NoStop}%
\bibitem [{\citenamefont {Aschwanden}(2011)}]{Asc11}%
  \BibitemOpen
  \bibfield  {author} {\bibinfo {author} {\bibfnamefont {M.}~\bibnamefont
  {Aschwanden}},\ }\href@noop {} {\emph {\bibinfo {title} {Self-Organized
  Criticallity in Astrophysics: Statistics of Nonlinear Processes in the
  Universe}}}\ (\bibinfo  {publisher} {Springer-Verlag},\ \bibinfo {address}
  {Berlin, Heidelberg},\ \bibinfo {year} {2011})\BibitemShut {NoStop}%
\bibitem [{\citenamefont {Gutenberg}\ and\ \citenamefont
  {Richter}(1955)}]{Gut55}%
  \BibitemOpen
  \bibfield  {author} {\bibinfo {author} {\bibfnamefont {B.}~\bibnamefont
  {Gutenberg}}\ and\ \bibinfo {author} {\bibfnamefont {C.~F.}\ \bibnamefont
  {Richter}},\ }\href@noop {} {\bibfield  {journal} {\bibinfo  {journal}
  {Nature}\ }\textbf {\bibinfo {volume} {176}},\ \bibinfo {pages} {795}
  (\bibinfo {year} {1955})}\BibitemShut {NoStop}%
\bibitem [{\citenamefont {Beggs}\ and\ \citenamefont {Plenz}(2003)}]{Beg03}%
  \BibitemOpen
  \bibfield  {author} {\bibinfo {author} {\bibfnamefont {J.~M.}\ \bibnamefont
  {Beggs}}\ and\ \bibinfo {author} {\bibfnamefont {D.}~\bibnamefont {Plenz}},\
  }\href@noop {} {\bibfield  {journal} {\bibinfo  {journal} {The Journal of
  Neuroscience}\ }\textbf {\bibinfo {volume} {23}},\ \bibinfo {pages} {11167}
  (\bibinfo {year} {2003})}\BibitemShut {NoStop}%
\bibitem [{\citenamefont {Witten}(1982)}]{Witten_SUSY}%
  \BibitemOpen
  \bibfield  {author} {\bibinfo {author} {\bibfnamefont {E.}~\bibnamefont
  {Witten}},\ }\href@noop {} {\bibfield  {journal} {\bibinfo  {journal}
  {Journal of Differential Geometry}\ }\textbf {\bibinfo {volume} {17}},\
  \bibinfo {pages} {661} (\bibinfo {year} {1982})}\BibitemShut {NoStop}%
\bibitem [{\citenamefont {Frenkel}\ \emph {et~al.}(2007)\citenamefont
  {Frenkel}, \citenamefont {Losev},\ and\ \citenamefont
  {Nekrasov}}]{TFT_Frankel}%
  \BibitemOpen
  \bibfield  {author} {\bibinfo {author} {\bibfnamefont {E.}~\bibnamefont
  {Frenkel}}, \bibinfo {author} {\bibfnamefont {A.}~\bibnamefont {Losev}}, \
  and\ \bibinfo {author} {\bibfnamefont {N.}~\bibnamefont {Nekrasov}},\
  }\href@noop {} {\bibfield  {journal} {\bibinfo  {journal} {Nuclear Physics B
  - Proceedings Supplements}\ }\textbf {\bibinfo {volume} {171}},\ \bibinfo
  {pages} {215 } (\bibinfo {year} {2007})},\ \bibinfo {note} {the proceedings
  of the International Conference on Strings and Branes: The present paradigm
  for gauge interactions and cosmology (Carg{\`e}se School on String Theory)
  Carg{\`e}se 2006}\BibitemShut {NoStop}%
\bibitem [{\citenamefont {Labastida}(1989)}]{TFT_Labastida}%
  \BibitemOpen
  \bibfield  {author} {\bibinfo {author} {\bibfnamefont {J.~M.~F.}\
  \bibnamefont {Labastida}},\ }\href@noop {} {\bibfield  {journal} {\bibinfo
  {journal} {Communications in Mathematical Physics}\ }\textbf {\bibinfo
  {volume} {123}},\ \bibinfo {pages} {641} (\bibinfo {year}
  {1989})}\BibitemShut {NoStop}%
\bibitem [{\citenamefont {Witten}(1988{\natexlab{a}})}]{TFT_Witten_Sigma}%
  \BibitemOpen
  \bibfield  {author} {\bibinfo {author} {\bibfnamefont {E.}~\bibnamefont
  {Witten}},\ }\href {\doibase 10.1007/BF01466725} {\bibfield  {journal}
  {\bibinfo  {journal} {Communications in Mathematical Physics}\ }\textbf
  {\bibinfo {volume} {118}},\ \bibinfo {pages} {411} (\bibinfo {year}
  {1988}{\natexlab{a}})}\BibitemShut {NoStop}%
\bibitem [{\citenamefont {Witten}(1988{\natexlab{b}})}]{TFT_Witten}%
  \BibitemOpen
  \bibfield  {author} {\bibinfo {author} {\bibfnamefont {E.}~\bibnamefont
  {Witten}},\ }\href {\doibase 10.1007/BF01223371} {\bibfield  {journal}
  {\bibinfo  {journal} {Communications in Mathematical Physics}\ }\textbf
  {\bibinfo {volume} {117}},\ \bibinfo {pages} {353} (\bibinfo {year}
  {1988}{\natexlab{b}})}\BibitemShut {NoStop}%
\bibitem [{\citenamefont {Birmingham}\ \emph {et~al.}(1991)\citenamefont
  {Birmingham}, \citenamefont {Blau}, \citenamefont {Rakowski},\ and\
  \citenamefont {Thompson}}]{TFT_Review}%
  \BibitemOpen
  \bibfield  {author} {\bibinfo {author} {\bibfnamefont {D.}~\bibnamefont
  {Birmingham}}, \bibinfo {author} {\bibfnamefont {M.}~\bibnamefont {Blau}},
  \bibinfo {author} {\bibfnamefont {M.}~\bibnamefont {Rakowski}}, \ and\
  \bibinfo {author} {\bibfnamefont {G.}~\bibnamefont {Thompson}},\ }\href@noop
  {} {\bibfield  {journal} {\bibinfo  {journal} {Physics Reports}\ }\textbf
  {\bibinfo {volume} {209}},\ \bibinfo {pages} {129 } (\bibinfo {year}
  {1991})}\BibitemShut {NoStop}%
\bibitem [{\citenamefont {Mostafazadeh}(2002)}]{pseudo_Hermitian}%
  \BibitemOpen
  \bibfield  {author} {\bibinfo {author} {\bibfnamefont {A.}~\bibnamefont
  {Mostafazadeh}},\ }\href@noop {} {\bibfield  {journal} {\bibinfo  {journal}
  {Nuclear Physics B}\ }\textbf {\bibinfo {volume} {640}},\ \bibinfo {pages}
  {419 } (\bibinfo {year} {2002})}\BibitemShut {NoStop}%
\bibitem [{\citenamefont {Ruelle}(2014)}]{Ruelle_Chaos}%
  \BibitemOpen
  \bibfield  {author} {\bibinfo {author} {\bibfnamefont {D.}~\bibnamefont
  {Ruelle}},\ }\href@noop {} {\bibfield  {journal} {\bibinfo  {journal}
  {Physics Today}\ }\textbf {\bibinfo {volume} {67}},\ \bibinfo {pages} {9}
  (\bibinfo {year} {2014})}\BibitemShut {NoStop}%
\bibitem [{\citenamefont {Manning}(2006)}]{BookEntropy}%
  \BibitemOpen
  \bibfield  {author} {\bibinfo {author} {\bibfnamefont {A.}~\bibnamefont
  {Manning}},\ }in\ \href@noop {} {\emph {\bibinfo {booktitle} {Dynamical
  Systems}}},\ \bibinfo {series} {Lecture Notes in Mathematics}, Vol.\ \bibinfo
  {volume} {468}\ (\bibinfo  {publisher} {Springer},\ \bibinfo {year} {2006})\
  p.\ \bibinfo {pages} {185}\BibitemShut {NoStop}%
\bibitem [{\citenamefont {Gilmore}(1998)}]{Gilmore}%
  \BibitemOpen
  \bibfield  {author} {\bibinfo {author} {\bibfnamefont {R.}~\bibnamefont
  {Gilmore}},\ }\href@noop {} {\bibfield  {journal} {\bibinfo  {journal} {Rev.
  Mod. Phys.}\ }\textbf {\bibinfo {volume} {70}},\ \bibinfo {pages} {1455}
  (\bibinfo {year} {1998})}\BibitemShut {NoStop}%
\bibitem [{\citenamefont {Ruelle}(2002)}]{Rue02}%
  \BibitemOpen
  \bibfield  {author} {\bibinfo {author} {\bibfnamefont {D.}~\bibnamefont
  {Ruelle}},\ }\href@noop {} {\bibfield  {journal} {\bibinfo  {journal}
  {Notices of AMS}\ }\textbf {\bibinfo {volume} {49}},\ \bibinfo {pages} {887}
  (\bibinfo {year} {2002})}\BibitemShut {NoStop}%
\end{thebibliography}%

\end{document}